\documentclass[pre,floats,twocolumn,showpacs,floatfix,superscriptaddress]{revtex4}
\usepackage[latin1]{inputenc}
\usepackage{graphicx}
\usepackage{amsmath}
\usepackage{amssymb}
\usepackage{pifont}
\usepackage[dvips]{color}
\usepackage{subeqnarray}
\usepackage{psfrag}
\begin{document}
\title{Tensorial slip of super-hydrophobic channels}
\author{Sebastian Schmieschek}
\affiliation{Institute for Computational Physics, University of Stuttgart, Pfaffenwaldring 27, D-70569 Stuttgart, Germany}
\affiliation{Department of Applied Physics, Eindhoven University of Technology, P.O. Box 513, 5600 MB Eindhoven, The Netherlands}
\author{Aleksey V. Belyaev}
\affiliation{Department of Physics, M.V.~Lomonosov Moscow State University, 119991 Moscow, Russia }
\affiliation{A.N.~Frumkin Institute of Physical
Chemistry and Electrochemistry, Russian Academy of Sciences, 31
Leninsky Prospect, 119991 Moscow, Russia}
\author{Jens Harting}
\affiliation{Department of Applied Physics, Eindhoven University of Technology, P.O. Box 513, 5600 MB Eindhoven, The Netherlands}
\affiliation{Institute for Computational Physics, University of Stuttgart, Pfaffenwaldring 27, D-70569 Stuttgart, Germany}
\author{Olga I. Vinogradova}
\affiliation{Department of Physics, M.V.~Lomonosov Moscow State University, 119991 Moscow, Russia }
\affiliation{A.N.~Frumkin Institute of Physical
Chemistry and Electrochemistry, Russian Academy of Sciences, 31
Leninsky Prospect, 119991 Moscow, Russia}
\affiliation{DWI, RWTH Aachen, Forckenbeckstr. 50, 52056 Aachen, Germany}
\date{\today}

\begin{abstract}
We describe a generalization of the tensorial slip boundary condition,
originally justified for a thick (compared to texture period) channel, to
any channel thickness. The eigenvalues of the effective slip length
tensor, however, in general case become dependent on the gap and cannot be
viewed as a local property of the surface, being a global characteristic
of the channel. To illustrate the use of the tensor formalism we develop a
semi-analytical theory of an effective slip in a parallel-plate channel
with one super-hydrophobic striped and one hydrophilic surface. Our
approach is valid for any local slip at the gas sectors and an arbitrary
distance between the plates, ranging from a thick to a thin channel.  We
then present results of lattice Boltzmann simulations to validate the
analysis. Our results may be useful for extracting effective slip tensors
from global measurements, such as the permeability of a channel, in
experiments or simulations.
\end{abstract}


\maketitle

\section{Introduction}
With recent advances in microfluidics~\cite{stone2004,squires2005}, renewed
interest has emerged in quantifying the effects of surface heterogeneities
with different local hydrophobicity (characterized by a local scalar
slip~\cite{vinogradova1999,lauga2005}), on fluid motion. In this situation it
is advantageous to construct the effective slip boundary condition, which is
applied at the hypothetical smooth homogeneously slipping surface, and mimics
the actual one along the true heterogeneous
surface~\cite{vinogradova.oi:2010,kamrin.k:2010}. Such an effective condition
fully characterizes the flow at the real surface and can be used to solve
complex hydrodynamic problems without tedious calculations. Well-known
examples of such a heterogeneous system include super-hydrophobic Cassie (SH)
surfaces, where trapped gas is stabilized with a rough wall texture, leading
to a number of `super' properties, such as extreme non-wettability and low
hysteresis~\cite{quere.d:2005}. For these surfaces effective slip lengths are
often very
large~\cite{choi.ch:2006,joseph.p:2006,tsai.p:2009,doi:10.1146/annurev-fluid-121108-145558}
compared a smooth hydrophobic
coating~\cite{vinogradova.oi:2009,vinogradova.oi:2003,bib:jens-jari:2008,bib:jens-kunert-herrmann:2005,charlaix.e:2005,PhysRevLett.96.046101},
which can greatly reduce the viscous drag and impact transport phenomena in
microchannels~\cite{vinogradova.oi:2010}.

\begin{figure}
\includegraphics[width=0.3\textwidth]{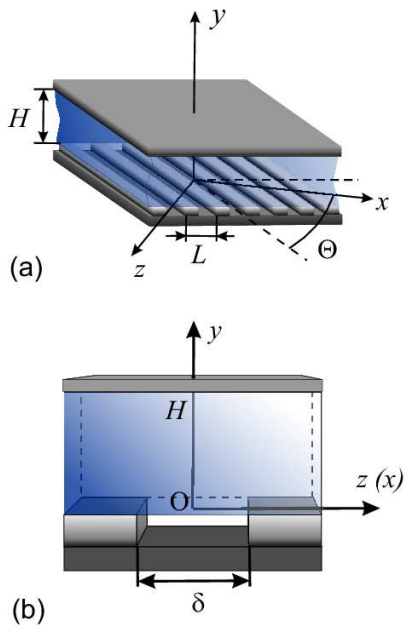}
\caption{Sketch of the SH stripes (a): $\Theta=\pi/2$ corresponds to
  transverse stripes, whereas $\Theta=0$ to longitudinal stripes; (b) situation
in (a) is approximated by a periodic cell of size $L$, with equivalent flow
boundary conditions on the gas-liquid and solid-liquid interfaces.
}\label{fig:geometry}
\end{figure}

The concept of effective slip was mostly exploited for \emph{thick} (compared
to the texture characteristic length, $L$)
channels~\cite{stone2004,ybert.c:2007}, where for an anisotropic texture it was
shown to depend on the direction of the flow and is a tensor~\cite{Bazant08},
$\textbf{b}_{\rm eff}\equiv\{b^{\rm eff}_{ij}\}$, represented by a symmetric,
positive definite $2\times 2$ matrix
\begin{equation} \label{beff_def1}
\mathbf{b}_{\rm eff} = \mathbf{S}_{\Theta}\left(
  \begin{array}{cc}
    b^{\parallel}_{\rm eff} & 0 \\
    0 & b^{\perp}_{\rm eff} \\
  \end{array}\right) \mathbf{S}_{-\Theta},
\end{equation}
diagonalized by a rotation with angle $\Theta$
\begin{equation} \label{S_def1}
\mathbf{S}_{\Theta}=\left(
  \begin{array}{cc}
\cos \Theta & \sin \Theta \\
- \sin \Theta & \cos \Theta \\
  \end{array}
\right).
\end{equation}
For all anisotropic surfaces its eigenvalues $b^{\parallel}_{\rm eff}$ and
$b^{\perp}_{\rm eff}$ correspond to the fastest (greatest forward slip) and
slowest (least forward slip) orthogonal directions~\cite{Bazant08}. In the
general case of any direction $\Theta$, this means that the flow past such
surfaces becomes misaligned with the driving force. This tensorial slip
approach, based on a consideration of a `macroscale' fluid motion instead of
solving hydrodynamic equations at the scale of the individual pattern, was
supported by statistical diffusion arguments~\cite{Bazant08}, and was
recently justified for the case of Stokes flow over a broad class of periodic
surfaces~\cite{kamrin.k:2010}. Note that an effective slip in a thick channel
situation is a characteristics of a heterogeneous interface solely (being
expressed through its parameters, such as local slip lengths, fractions of
phases, and a texture period)~\cite{vinogradova.oi:2010,bocquet2007}.

It was however recently recognized and justified by using the theory of
heterogeneous porous materials~\cite{feuillebois.f:2009} that a similar
concept of effective slip can be also exploited for a flow conducted in a
\emph{thin} channel with two confining surfaces separated by a distance $H
\ll L$. In such a situation a natural definition of the effective slip length
could be based on a permeability of a hypothetical uniform channel with the
same flow rate. An effective tensorial slip is then determined by flow at
the scale of the channel width, and depends on $H$~\cite{feuillebois.f:2009}.
This points to the fact that an effective boundary condition reflects not
just parameters of the liquid-solid interface, but also depends on the flow
configuration~\cite{vinogradova.oi:2010}.

The power of the effective slip approach and the super-lubrication potential of
SH surfaces  were already illustrated by discussing several applications. In
particular, it has been shown that optimized SH textures may be successfully
used in passive microfluidic
mixing~\cite{feuillebois.f:2010b,vinogradova.oi:2010} and for a reduction of a
hydrodynamic drag force~\cite{belyaev.av:2010b,asmolov.es:2011}, and that
the effective slip formalism represents a useful tool to quantify properties of
SH surfaces in thick and thin channels. In many situations however the dramatic
changes in flow happen when the two length scales $H$ and $L$ are of the same
order. In this work we generalize the definition of the effective slip length
tensor Eq.~(\ref{beff_def1}) to an \emph{arbitrary} channel thickness and
validate this approach by means of lattice Boltzmann (LB) simulations.

The structure of this manuscript is as follows. In section II we describe our
model and formulate governing equations. In section III we analyze a
tensorial permeability of the parallel-plate channel with one (anisotropic)
SH and one hydrophilic surface and give some general arguments showing that
the tensorial relation, Eq.~(\ref{beff_def1}), should be valid at  arbitrary
$H/L$. Section IV contains semi-analytical results for a striped SH surface,
and in section V we describe our LB simulation approach. In section VI
simulation results are presented to validate the predictions of the tensorial
theory and to test our analytical results for the asymptotic limits of thick
and thin channels.

\section{Model and governing equations}
The basic assumptions of our theoretical model are as follows. We consider a
channel consisting of two parallel walls located at $y=0$ and $y=H$ and
unbounded in the $x$ and $z$ directions as sketched in
Fig.\ref{fig:geometry}. The upper plate represents a no-slip hydrophilic
surface, and the lower plate is a SH surface. The origin of coordinates is
placed it the plane of a liquid-gas interface at the center of the gas
sector. The $x$ axis is defined along the pressure gradient. This (SH vs.
hydrophilic) geometry of configuration is relevant for various setups, where
the alignment of opposite textures is inconvenient or difficult. Besides that
the advantage of such a geometry is that it allows to avoid the gas bridging
and long-range attractive capillary forces~\cite{andrienko.d:2004}, which
appear when we deal with interactions of two hydrophobic
solids~\cite{yakubov:00,tyrrell.jwg:2001}.

As in previous
publications~\cite{bocquet2007,belyaev.av:2010b,priezjev.nv:2005}, we model
the SH plate as a flat interface with no meniscus curvature, so that the SH
surface appears as perfectly smooth with a pattern of boundary conditions.
The latter are taken as no-slip ($b_1=0$) over solid/liquid areas and as
partial slip ($b_2=b$) over gas/liquid regions. We denote as $\delta$ a
typical length scale of gas/liquid areas. The fraction of the solid/liquid areas is denoted $\phi_1=(L-\delta)/L$, and of the gas/liquid areas
$\phi_2=1-\phi_1=\delta/L$.  In this idealization, by assuming a flat
interface, we neglect an additional mechanism for a dissipation
connected with the meniscus
curvature~\cite{bib:jens-jari:2008,lauga2009,sbragaglia.m:2007}.

The flow is governed by the Stokes equations
\begin{equation}\label{NS}
   \eta\nabla^2\textbf{u}=\nabla p,
\end{equation}
\begin{equation}\label{NC}
  \nabla\cdot\textbf{u}=0,
\end{equation}
where $\textbf{u}$ is the velocity vector, and the average pressure gradient is always aligned with the $x$-axis direction:
\begin{equation}\label{PressGrad}
   \langle\nabla p\rangle = (-\sigma, 0, 0)
\end{equation}
The local slip boundary conditions at the walls are defined as
\begin{equation}\label{BC1}
   {\bf u}(x,0,z) = b(x,z)\cdot\frac{\partial {\bf u}}{\partial y}(x,0,z), \quad\hat{{\bf y}}\cdot {\bf u}(x,0,z) = 0,
\end{equation}
\begin{equation}\label{BC2}
   \textbf{u}(x,H,z)=0,\quad\hat{{\bf y}}\cdot {\bf u}(x,H,z) = 0.
\end{equation}
Here the local slip length $b(x,z)$ at the SH surface is generally the
function of both lateral coordinates.

We intend to evaluate the effective slip length $b_{\rm eff}$ at the SH
surface, which is as usual defined as
\begin{equation} \label{definition}
   b_{\rm eff}=\frac{\langle u_s\rangle}{\left\langle \left(\frac{\partial u}{\partial y}\right)_s \right\rangle},
\end{equation}
where $\langle \ldots\rangle$ means the average value in the plane $xOz$.

\section{General consideration}
In this section, we evaluate the pressure-driven flow in the parallel-plate
channel with one anisotropic SH and one hydrophilic surface and investigate the
consequences of anisotropy. For an anisotropic texture, there are several
possible ways to define effective slip lengths. A natural approach is to define
a slip-length tensor via
\begin{equation}\label{k_b}
   k^{\parallel, \perp}= \frac{H^3}{12}\left(1+\frac{3 b_{\rm eff}^{\parallel, \perp}}{H+b_{\rm eff}^{\parallel, \perp}}\right)
\end{equation}
by analogy with a hypothetical uniform channel. Such a definition was earlier
justified for thin SH channels, by using the lubrication
limit~\cite{feuillebois.f:2009}. Below we argue that the same result is
obtained for a channel of an arbitrary thickness with a tensorial
generalization of the Navier boundary condition, where $\textbf{b}_{\rm eff}$
is a global measure of the effective slippage of the channel.

For a mathematical justification of the above statement we first rewrite
Eqs.~(\ref{NS}) and (\ref{PressGrad}) as
\begin{equation}
   \eta \nabla^2 \langle\textbf{U}\rangle = \langle \nabla p \rangle, \qquad \langle \nabla p \rangle= -\sigma \hat{\textbf{x}},
\end{equation}
where $\langle\textbf{U}\rangle$ is the depth-averaged velocity. The two boundary conditions that apply at the channel walls can then be written as follows: $\langle\textbf{U}\rangle=0$ at the upper surface, and uniform tensorial slip,
\begin{equation}
   \langle\textbf{U}\rangle = \mathbf{b}_{\rm eff} \cdot \langle \partial_y \textbf{U}\rangle,
\end{equation}
at the lower surface, with $\mathbf{b}_{\rm eff}$ defined according to
Eqs.~(\ref{beff_def1})-(\ref{S_def1}).  The solution for the depth-averaged
velocity is then given by~\cite{vinogradova.oi:2010}
\begin{eqnarray}
   \langle U_x\rangle &=& -\frac{\sigma y^2}{2\eta}+\frac{\sigma H y}{2\eta}+\frac{\sigma H^2}{2\eta}C_x\left(1-\frac{y}{H}\right), \\
   \langle U_z\rangle &=& \frac{\sigma H^2}{2\eta} C_z\left(1-\frac{y}{H}\right),
\end{eqnarray}
with
\[
  C_x=  \frac{H b_{\rm eff}^{\parallel} \cos^2\Theta+H b_{\rm eff}^{\perp} \sin^2\Theta+b_{\rm eff}^{\parallel}b_{\rm eff}^{\perp}}{(H+b_{\rm eff}^{\parallel})(H+b_{\rm eff}^{\perp})},\]
\[  C_z=\frac{H(b_{\rm eff}^{\parallel}-b_{\rm eff}^{\perp})\sin\Theta\cos\Theta}{(H+b_{\rm eff}^{\parallel})(H+b_{\rm eff}^{\perp})}.
\]

In linear response, the averaged flow rate, $\langle\textbf{Q}\rangle$, is
proportional to $\langle \nabla p \rangle$ via the permeability tensor,
$\textbf{k}$:
\begin{equation}\label{Darcy_tensor}
    \langle\textbf{Q}\rangle = -\frac{1}{\eta} \textbf{k}\cdot \langle \nabla p \rangle.
\end{equation}
Integrating the velocity profile across the channel we obtain
\begin{equation}\label{Q}
    \langle\textbf{Q}\rangle=\int\limits_0^H{\left\langle\textbf{U}(y)\right\rangle dy},
\end{equation}
with the components
\begin{equation}\label{Q_I}
    \langle Q \rangle_x =\frac{\sigma}{\eta} \frac{H^3}{12} \left[1+3 C_x \right], \, \,
    \langle Q \rangle_z = \frac{\sigma}{\eta} \frac{H^3}{4} C_z.
\end{equation}
The latter may be rewritten as
\begin{eqnarray}\label{Q_II}
    \langle Q \rangle_x &=&\frac{\sigma}{\eta} \left(k^{\parallel}\cos^2\Theta+k^{\perp}\sin^2\Theta\right),\\
    \langle Q \rangle_z &=&\frac{\sigma}{\eta}\left(k^{\parallel}-k^{\perp}\right)\sin\Theta\cos\Theta,
\end{eqnarray}
provided the two tensors, $\textbf{k}$ and $\textbf{b}_{\rm eff}$, are coaxial
and the rigorous relationship between their eigenvalues is given by
Eq.~(\ref{k_b}). This unambiguously indicates that the two definitions of the
slip length are equivalent. It becomes also apparent that Eq.~(\ref{k_b})
implies that $b_{\rm eff}^{\parallel, \perp}$ generally depends on the separation $H$.

Note some similarity to a prior
work~\cite{feuillebois.f:2009,strook_et_al_2:2002}. The current consideration,
however, is valid for arbitrary $H/L$ and $b/L$, including the limit of thick
channels, where $\textbf{b}_{\rm eff}$ becomes a local property of the surface
on scales much larger than the texture characteristic length.

Consider now a situation where the `fast' axis of greatest forward slip of
anisotropic texture is inclined at an angle $\Theta$ to the pressure gradient.
This problem can be solved explicitly as follows. The downstream effective
permeability of the channel can be expressed in terms of the effective
downstream slip length as
\begin{equation}\label{downstream_b}
  k^{(x)}_{\rm eff} =  \frac{H^3}{12}\left(1+\frac{3 b_{\rm eff}^{(x)}}{H+b_{\rm eff}^{(x)}}\right).
\end{equation}
Following~\cite{Bazant08}, it can also be obtained
from the permeability tensor:
\begin{equation}\label{downstream_k}
  k^{(x)}_{\rm eff} =  \frac{k^{\parallel} k^{\perp}}{k^{\parallel} \sin^2{\Theta} + k^{\perp}\cos^2{\Theta}}
\end{equation}
By substituting Eq.~(\ref{k_b}) into Eq.~(\ref{downstream_k}) and after
subtracting the latter from Eq.~(\ref{downstream_b})
we express an effective downstream slip length in the form
\begin{eqnarray}\label{b_eff_x_1}
   b_{\rm eff}^{(x)}= \frac{b_{\rm eff}^{ \perp}H+4b_{\rm eff}^{ \parallel}b_{\rm eff}^{ \perp}+(b_{\rm eff}^{ \parallel}-b_{\rm eff}^{ \perp})H\cos^2\Theta}{H+4b_{\rm eff}^{ \parallel}-4(b_{\rm eff}^{ \parallel}-b_{\rm eff}^{ \perp})\cos^2\Theta}
\end{eqnarray}
Note that in the general case $b_{\rm eff}^{(x)}$ depends on $H$ and $b_{\rm
eff}^{ \parallel, \perp}(H)$. For this reason, $ b_{\rm eff}^{(x)}$ cannot be
viewed as a \emph{local} property of the SH surface, except as in the thick
channel limit. Instead, it is generally the effective slip length of the SH
channel and thus its \emph{global} characteristic.

Finally, we emphasize the generality of Eqs.~(\ref{downstream_k}) and
(\ref{b_eff_x_1}), which follow only from the symmetry of the effective
permeability and slip- length tensors for linear response. Similar formulae
have been obtained before in a few particular
calculations~\cite{strook_et_al_2:2002, ajdariPRE2001}, but the present derivation
is valid regardless of the thickness of the channel and is independent of the
details of the textured surface. There could be arbitrary patterns of local
slip lengths, and the latter could itself be a spatially varying tensor,
reflecting surface anisotropy at a smaller (possibly atomic) scales.

\section{Theory for striped patterns}
\label{fourier}
To illustrate the general theory, in this section we focus on flat patterned
SH surfaces consisting of periodic stripes, where the local (scalar) slip
length $b$ varies only in one direction. The problem of flow past striped SH
surfaces has previously been studied in the context of a reduction of
pressure-driven forward flow in
thick~\cite{lauga.e:2003,belyaev.av:2010a,priezjev.nv:2005} and
thin~\cite{feuillebois.f:2009} channels, and it is directly relevant for
mixing~\cite{feuillebois.f:2010b,vinogradova.oi:2010}, and a generation of a
tensorial electro-osmotic
flow~\cite{bahga:2010,vinogradova.oi:2010,belyaev.av:2011a}. Here we elaborate
on a previously published anzatz~\cite{vinogradova.oi:2010} and present the
theory for an arbitrary gap, which in the asymptotic limits describes
situations of thin and thick channels. The mathematical analysis we use here
is similar to a technique exploited in~\cite{belyaev.av:2010a} for a thick
channel configuration. The crucial difference with
Ref.~\cite{belyaev.av:2010a}, however, is that we consider an arbitrary gap,
which means that the effective slip is a function of the channel thickness as
discussed above.

For transverse stripes, we have ${\bf u}=(u(x,y), v(x,y), 0)$,
$u(x,0)=b(x)u_y(x,0)$, and $v(x,0)=0$. For longitudinal stripes, the flow is
also two dimensional: ${\bf u}=(u(y,z), v(y,z),0)$, $u(0,z)=b(z)u_y(0,z)$, and
$v(0,z)=0$.  As the problem is linear in \textbf{u}, we seek a solution of the
form
\begin{equation}
   \textbf{u}=\textbf{u}_0+\textbf{u}_1,
\end{equation}
where $\textbf{u}_0$ is the velocity of the usual no-slip parabolic Poiseuille
flow
\begin{equation}
   \textbf{u}_0=(u_0, 0, 0),\qquad u_0=-\frac{\sigma}{2\eta} y^2 + \frac{\sigma H}{2\eta} y
\end{equation}
and $\textbf{u}_1$ is the SH slip-driven superimposed flow.

\subsection{Longitudinal stripes}
In this situation the problem is homogeneous in $x$-direction
($\partial/\partial x=0$). The slip length $b(x,z)=b(z)$ is periodic in $z$
with period $L$. The elementary cell is determined as $b(z)=b$ at
$|z|\leq\delta/2$, and $b(z)=0$ at $\delta/2<|z|\leq L/2$.  In this case the
velocity $\textbf{u}_1=(u_1, 0, 0)$ has only one nonzero component, which can
be determined by solving the Laplace equation
\begin{equation}\label{BP_u1_longit}
   \nabla^2 u_1(y,z)=0,
   \end{equation}
   with the following boundary conditions defined in the usual way as
   \begin{equation}\label{BC2_long}
   u_1(H,z)=0,
\end{equation}
   \begin{equation}\label{BC1_long}
   u_1(0,z)=b(z)\left(\frac{\sigma H}{2\eta}+\left.\frac{\partial u_1}{\partial y}\right|_{y=0}\right).
   \end{equation}
The Fourier method yields a general solution to this problem:
\begin{eqnarray}\label{u1longit}
   u_1(y,z)&=&(M_0y+P_0)\\&+&\sum^\infty_{n=1} (M_n e^{\lambda_n y}+ P_n e^{-\lambda_n y}) \cos(\lambda_n z),\nonumber
\end{eqnarray}
with $\lambda_n=2\pi n/L$. The sine terms vanish due to symmetry.
Condition (\ref{BC2_long}) leads to
\begin{eqnarray}\label{u1longit_2}
   u_1(y,z)&=&P_0\left(1-\frac{y}{H}\right)\\&+&\sum^\infty_{n=1}P_n\cos(\lambda_n z) e^{-\lambda_n y} (1- e^{-2\lambda_n (H-y)}).\nonumber
\end{eqnarray}
Applying the boundary condition (\ref{BC1_long}) we then obtain a trigonometric
dual series:
\begin{eqnarray}\label{DoubleSeries_a}
   a_0\left(1+\frac{b}{H}\right)&+&\sum^\infty_{n=1} a_n \left[1+b \lambda_n \coth(\lambda_n H) \right] \cos(\lambda_n z)\nonumber\\&=&b \frac{\sigma H}{2\eta} ,\quad 0<z\leq \delta/2,   
\end{eqnarray}
\begin{equation}\label{DoubleSeries_b}
   a_0+\sum^\infty_{n=1} a_n \cos(\lambda_n z)=0,\quad \delta/2<z\leq L/2,
\end{equation}
where
\[
   a_0=P_0; \quad a_n=P_n(1-e^{-2\lambda_nH}),\: n\geq 1.
\]
Dual series (\ref{DoubleSeries_a}), (\ref{DoubleSeries_b}) provide a complete
description of hydrodynamic flow and effective slip in the longitudinal
direction, given all the stated assumptions. These equations can be solved
numerically (see Appendix~\ref{A1}), but exact results are possible in the limits of
thin and thick channels.

For a thin channel, $H\ll L$, we can use that $\left. \coth t
\right|_{t\rightarrow0}=t^{-1}+O(t)$. By substituting this expression into
(\ref{DoubleSeries_a}) and and keeping only values of the first non-vanishing order \cite{notice1}, we find
\begin{equation}\label{a0_longit}
   a_0=\frac{2}{L} \int\limits_0^{\delta/2} {\frac{\sigma H}{2\eta}\frac{b}{1+b/H}  dz}=\frac{\sigma H}{2\eta}\frac{b H \phi_2}{H+b}
\end{equation}
whence~\cite{feuillebois.f:2009}
\begin{equation}\label{beff_longit_smallH1}
   \left.b_{\rm eff}^{\parallel}\right|_{H\rightarrow 0} = \frac{b H \phi_2}{H + b \phi_1}.
\end{equation}
This is an \emph{exact} solution, representing a rigorous upper Wiener bound
on the effective slip over all possible two-phase patterns in a thin channel.
In order to gain a simple physical understanding of this result and to
facilitate the analysis below, it is instructive to mention the two
limits~\cite{belyaev.av:2010b,vinogradova.oi:2010} that follow from
Eq.~(\ref{beff_longit_smallH1}). When
$H\ll b, L$ we deduce
\begin{equation}\label{yy1}
  \left. b_{\rm eff}^{\parallel}\right|_{H\ll b, L} \simeq \frac{\phi_2}{\phi_1}H  \:  \propto H,
\end{equation}
and when $b\ll H\ll L$ we get a surface averaged slip
\begin{equation}\label{xx1}
  \left. b_{\rm eff}^{\parallel}\right|_{b\ll H\ll L} \simeq b\phi_2  \:  \propto b.
\end{equation}

In the limit of a thick channel, $H\gg L$, we can use that
$\coth(t\rightarrow \infty)\rightarrow 1$ and the dual series
(\ref{DoubleSeries_a})-(\ref{DoubleSeries_b}) can be solved exactly to
obtain\cite{vinogradova.oi:2010}
\begin{equation}\label{beff_par_largeH}
  b_{\rm eff}^{\parallel} \simeq \frac{L}{\pi} \frac{\ln\left[\sec\left(\displaystyle\frac{\pi \phi_2}{2 }\right)\right]}{1+\displaystyle\frac{L}{\pi b}\ln\left[\sec\displaystyle\left(\frac{\pi \phi_2}{2 }\right)+\tan\displaystyle\left(\frac{\pi \phi_2}{2}\right)\right]}.
\end{equation}
This expression for an effective slip length depends strongly on a texture
period $L$. When $b \ll L$ we again derive the area-averaged slip length
\begin{equation}\label{zz1}
  \left. b_{\rm eff}^{\parallel}\right|_{b\ll L\ll H} \simeq b\phi_2  \:  \propto b.
\end{equation}
When $b \gg L$, expression (\ref{beff_par_largeH}) takes
the form
\begin{equation}\label{beff_ort_largeH_id}
\left.  b_{\rm eff}^{\parallel} \right|_{L\ll b, H} \simeq \frac{L}{\pi} \ln\left[\sec\left(\displaystyle\frac{\pi \phi_2}{2 }\right)\right] \propto L,
\end{equation}
that coincides with an earlier result~\cite{lauga.e:2003} obtained  for a
perfect slip ($b\to \infty$) case.

\subsection {Transverse stripes}
In this case it is convenient to introduce a stream function $\psi(x,y)$ and
the vorticity vector $\boldsymbol{\omega} (x,y)$. The two-dimensional
velocity field corresponding to the transverse configuration is represented
by $\textbf{u}(x,y)=\left(\partial \psi/\partial y, -\partial \psi/\partial x
, 0 \right)$, and the vorticity vector, $ \boldsymbol {\omega}
(x,y)=\nabla\times\textbf{u}=(0,0,\omega)$, has only one nonzero component,
which equals to
\begin{equation}\label{vorticity}
   \omega=-\nabla^2\psi.
\end{equation}
The solution can then be presented as the sum of the base flow with homogeneous
no-slip condition and its perturbation caused by the presence of stripes as
\begin{equation}\label{SF01}
   \psi=\Psi_0+\psi_1,\quad \omega=\Omega_0+\omega_1,
\end{equation}
where $\Psi_0$ and $\Omega_0$ correspond to the typical Poiseuille flow in a
flat channel with no-slip walls:
\begin{equation}\label{Psi0}
   \Psi_0=-\frac{\sigma}{\eta}\frac{y^3}{6}+\frac{\sigma H}{\eta}\frac{y^2}{4},\quad \Omega_0=\frac{\sigma}{\eta}y-\frac{\sigma H}{2\eta}.
\end{equation}

The problem for perturbations of the stream function and $z$-component of the
vorticity vector reads
\begin{equation}\label{eq_psi1}
   \nabla^2\psi_1=-\omega_1,\quad \nabla^2\omega_1=0,
\end{equation}
which can be solved by applying boundary conditions
\begin{eqnarray}  \label{BC1_new}
   \frac{\partial \psi_1}{\partial y}(x, 0)&=& b(x)\cdot\left[\frac{\sigma H}{2\eta}-\omega_1(x, 0)\right],\nonumber\\
   \frac{\partial \psi_1}{\partial y}(x, H)&=& 0,\quad \frac{\partial \psi_1}{\partial x}(x, H)= 0
\end{eqnarray}
and an extra condition that reflects our definition of the stream function:
\begin{equation}
\psi_1(x,0) = 0
\end{equation}
This can be solved to get
\begin{eqnarray}\label{GenSol_psi1}
   &&\psi_1(x,y)=-\frac{M_0}{4}y^2+P_0 y\nonumber\\ &+&
   \sum^\infty_{n=1} \left(P^{(1)}_n - \frac{M^{(1)}_n}{2} \frac{y}{\lambda_n}\right)e^{\lambda_n y}\cos{\lambda_n x} \nonumber\\ &+&
   \sum^\infty_{n=1}\left(P^{(2)}_n + \frac{M^{(2)}_n}{2} \frac{y}{\lambda_n}\right)e^{-\lambda_n y}\cos{\lambda_n x},    
\end{eqnarray}
\begin{equation}\label{GenSol_omega1}
   \omega_1(x,y) = \frac{M_0}{2}+\sum^\infty_{n=1} \left(M^{(1)}_n e^{\lambda_n y} +M^{(2)}_n e^{-\lambda_n y}\right)\cos(\lambda_n x).
\end{equation}
Conditions (\ref{BC1_new}) lead to
\[
    P^{(1)}_n=-P^{(2)}_n\equiv -P_n,\quad M_0=\frac{2 P_0}{H},
\]
\[
    M^{(1)}_n=-\frac{P_n\left(-e^{\lambda_nH}+e^{-\lambda_nH}+2\lambda_nH e^{\lambda_nH}\right)}{H^2 e^{\lambda_nH}},
\]
\[
    M^{(2)}_n=-\frac{P_n\left(-e^{\lambda_nH}+e^{-\lambda_nH}+2\lambda_nH e^{-\lambda_nH}\right)}{H^2 e^{-\lambda_nH}},
\]
and we obtain another dual series problem, which is similar to
(\ref{DoubleSeries_a}) and (\ref{DoubleSeries_b}):
\begin{eqnarray}\label{DoubleSeries_c}
      a_0\left(1+\frac{b}{H}\right)&+&\sum^\infty_{n=1} a_n \left[1+2 b \lambda_n V(\lambda_n H) \right] \cos(\lambda_n x)\nonumber\\&=&b \frac{\sigma H}{2\eta} ,\quad 0<x\leq \delta/2,   
\end{eqnarray}
\begin{equation}\label{DoubleSeries_d}
      a_0+\sum^\infty_{n=1} a_n \cos(\lambda_n x)=0,\quad \delta/2<x\leq L/2,
\end{equation}
Here, 
\[
   a_0=P_0; \quad
   a_n=\frac{\cosh(2 \lambda_n H)-2 \lambda_n^2 H^2 -1}{\lambda_n H^2}P_n, \: n\geq 1,
\]
and
\begin{equation}\label{V}
   V(t)=\frac{\sinh(2t)-2t}{\cosh(2t)-2t^2-1}.
\end{equation}

In the limit of a thin channel (where $V(t)|_{t\to \infty} \simeq
2t^{-1}+O(t)$), the dual series problem transforms to
\begin{eqnarray}\label{DoubleSeries_c_1}
   a_0+\left(1+\frac{3b}{H+b}\right)\cdot\sum^\infty_{n=1} a_n \cos(\lambda_n x)&=&\frac{\sigma H}{2\eta}\frac{b}{1+b/H} ,\quad \nonumber\\ && 0<x\leq \delta/2,
\end{eqnarray}
\begin{equation}\label{DoubleSeries_d_1}
   a_0+\sum^\infty_{n=1} a_n \cos(\lambda_n x)=0,\quad \delta/2 <x\leq L/2,
\end{equation}
which allows one to evaluate
\begin{equation}\label{a0_transv}
   a_0=\frac{\sigma H}{2\eta}\frac{b H \phi_2}{H+4b-3 \phi_2 b}.
\end{equation}
The effective slip length is then~\cite{feuillebois.f:2009}
\begin{equation}\label{beff_longit_smallH}
  \left. b_{\rm eff}^{\perp}\right|_{H\rightarrow 0} = \frac{b H \phi_2}{H + 4b \phi_1}.
\end{equation}
This \emph{exact} equation represents a rigorous lower Wiener bound on the effective slip over all possible two-phase patterns in a thin channel.

For completeness here we mention again the two limiting situations:
\begin{equation}\label{yy2}
   b_{\rm eff}^{\perp}|_{H\ll b, L} \simeq \frac{1}{4}\frac{\phi_2}{\phi_1}H  \:  \propto H,
\end{equation}
\begin{equation}\label{xx2}
   b_{\rm eff}^{\perp}|_{b\ll H\ll L} \simeq b\phi_2  \:  \propto b.
\end{equation}

In the thick channel limit, the dual series (\ref{DoubleSeries_c}) and
(\ref{DoubleSeries_d}) take the same form as in prior
work~\cite{vinogradova.oi:2010} (due to $V(x\rightarrow\infty)\rightarrow1$),
whence we derive~\cite{vinogradova.oi:2010}
\begin{equation}\label{beff_ort_largeH}
  b_{\rm eff}^{\perp} \simeq \frac{L}{2 \pi} \frac{\ln\left[\sec\left(\displaystyle\frac{\pi \phi_2}{2 }\right)\right]}{1+\displaystyle\frac{L}{2 \pi b}\ln\left[\sec\displaystyle\left(\frac{\pi \phi_2}{2 }\right)+\tan\displaystyle\left(\frac{\pi \phi_2}{2}\right)\right]}.
\end{equation}
The consideration as above of the same limits of small and large $b$ give
\begin{equation}\label{zz2}
  \left. b_{\rm eff}^{\perp}\right|_{b\ll L\ll H} \simeq b\phi_2  \:  \propto b
\end{equation}
and
\begin{equation}\label{beff_ort_largeH_id2}
\left.  b_{\rm eff}^{\perp} \right|_{L\ll b, H} \simeq \frac{L}{2\pi} \ln\left[\sec\left(\displaystyle\frac{\pi \phi_2}{2 }\right)\right] \propto L,
\end{equation}

\subsection{Tilted stripes}
If the stripes are inclined at an angle $\Theta$, the effective slip length
of the channel, $ b_{\rm eff}^{(x)}$ can be calculated with
Eq.~(\ref{b_eff_x_1}), provided that effective slip in eigendirections is
determined from the numerical solution of
(\ref{DoubleSeries_a}), (\ref{DoubleSeries_b}) and
(\ref{DoubleSeries_c}), (\ref{DoubleSeries_d}). Some simple analytical results
are possible in the limit of thin and thick channels.

In case of a thin channel and large local slip, $H\ll \min\{b,L\}$,
substitution of Eqs.~(\ref{yy1}) and (\ref{yy2}) into Eq. (\ref{b_eff_x_1})
gives
\begin{eqnarray}\label{b_eff_x_thin}
   b_{\rm eff}^{(x)} \simeq \frac{H\phi_2}{4\phi_1}\frac{4\phi_2+\phi_1+3\phi_1\cos^2\Theta}{4\phi_2+\phi_1-3\phi_2\cos^2\Theta}.
\end{eqnarray}
Interestingly, in this limit $b_{\rm eff}^{(x)}$ does not depend on $b$, being
a function of only $H$ and a fraction of the gas area. At small $b$  according
to Eqs.~(\ref{xx1}) and (\ref{xx2}), $b_{\rm eff}^{\parallel} \simeq b_{\rm
eff}^{\perp}\simeq  b_{\rm eff}^{(x)}$, so that the flow becomes isotropic.

In the limit of a thick channel and sufficiently large local slip, we can
simplify Eq.(\ref{b_eff_x_1}) and define the downstream effective slip length
as
\begin{eqnarray}\label{b_eff_x_thick}
   b_{\rm eff}^{(x)} \simeq \left(b_{\rm eff}^{ \parallel}-b_{\rm eff}^{ \perp}\right)\cos^2\Theta+b_{\rm eff}^{ \perp}
\end{eqnarray}
with  $b_{\rm eff}^{ \parallel, \perp}$ given by Eqs.(\ref{beff_par_largeH})
and (\ref{beff_ort_largeH}). In the limit of perfect local slip it can be
further simplified to get
\begin{equation}\label{simple_tensor}
     b_{\rm eff}^{(x)} \simeq b_{\rm eff}^{ \perp} (1+\cos^2\Theta).
\end{equation}
Using Eqs.~(\ref{zz1}) and (\ref{zz2}) we conclude that the flow is isotropic
$b_{\rm eff}^{\parallel} \simeq b_{\rm eff}^{\perp}\simeq  b_{\rm eff}^{(x)}$
when $b$ is much smaller than the texture period.

\section{Simulation method}\label{sec:lbm}
To simulate fluid flow between parallel patterned plates a number of
simulation methods could be used. These include Molecular Dynamics,
Dissipative Particle Dynamics, Stochastic Rotation Dynamics, classical
Finite Element or Finite Volume solvers as well the LB
method. Since the current paper does not address molecular interactions or
liquid-gas transitions close to the surface, one can limit the required
computational effort by applying a continuum solver for the Stokes
equation together with appropriate boundary conditions to model local
slip. As detailed further below, in particular in the thin channel limit
with $H \ll \min\{b,L\}$ a very high resolution of the flow field is
needed in order to measure $b_{\rm eff}$ with required precision.
Thus, simulation methods which require time averaging of the flow field or
finite lengths scales to resolve a slip boundary render less efficient for
the current problem. While Finite Element or Finite Volume solvers would
be suitable alternatives, we apply the LB
method~\cite{bib:succi-01}.

The LB approach is based on the Boltzmann kinetic equation
\begin{equation}
\label{eq:boltzmann}
\left[\frac{\partial }{\partial t}+{\bf u} \cdot \nabla_{\bf r}\right] f({\bf r,u},t)={\bf \Omega},
\end{equation}
which describes the evolution of the single particle probability density
$f({\bf r},{\bf u},t)$, where ${\bf r}$ is the position, ${\bf u}$ the
velocity, and $t$ the time. The derivatives on the left-hand side represent
propagation of particles in phase space whereas the collision operator ${\bf
\Omega}$ takes into account molecular collisions.

In the LB method the time $t$, the position ${\bf r}$, and the
velocity ${\bf u}$ are discretized. In units of the lattice constant $\Delta x$
and the time step $\Delta t$ this leads to a discretized version of
Eq.~(\ref{eq:boltzmann}):
\begin{equation}
\begin{array}{cc}
f_k({\bf r}+{\bf c}_k, t+1) - f_k({\bf r},t) =
\Omega_k, &  k= 0,1,\dots,B.
\end{array}
\end{equation}
Our simulations are performed on a three dimensional lattice with $B=19$
discrete velocities (the so-called D3Q19 model). With a proper choice of the
discretized collision operator ${\bf \Omega}$ it can be shown that the flow
behavior follows the Navier-Stokes equation~\cite{bib:succi-01}.  We choose the
Bhatnagar-Gross-Krook (BGK) form~\cite{bib:bgk}
\begin{equation}
\label{Omega}
 \Omega_k =
 -\frac{1}{\tau}\left(f_k({\bf r},t) - f_k^{eq}({\bf u}({\bf r},t),\rho({\bf r},t))\right)\mbox{ ,}
\end{equation}
which assumes a relaxation towards a discretized local Maxwell-Boltzmann
distribution $f_k^{eq}$. Here, $\tau$ is the mean collision time that
determines the kinematic viscosity $\nu=\frac{2\tau-1}{6}$ of the fluid. In
this study it is kept constant at $\tau=1.0$.

Physical properties of the simulated fluid are given by the stochastical
moments of the distribution function. Of special interest are the conserved
quantities, namely the fluid density  $\rho({\bf r},t) = \rho_{0} \sum_{k} f_{k}
(\mathbf{r}, t)$ and the momentum $\rho({\bf r},t) {\bf u}({\bf r},t) = \rho_{0} \sum_{k} c_{k}
f_{k} (\mathbf{r}, t)$, with $\rho_{0}$ being a reference density.

Within the LB method a common approach to describe the
interaction between hydrophobic surfaces and the fluid is by means of a
repulsive force
term~\cite{bib:jens-kunert-herrmann:2005,bib:jens-jari:2008,bib:zhu-tretheway-petzold-meinhart-2005,bib:benzi-etal-06,bib:zhang-kwok-04}.
This force applied at the boundary can be linked to the contact angle to
quantitatively describe the wettability of
materials~\cite{benzi-etal-06b,bib:huang-thorne-schaap-sukop-2007,bib:jens-schmieschek:2010,bib:jens-kunert-herrmann:2005}.
Alternatively, slip can be introduced by generalizing the no-slip bounce-back
boundary conditions in order to allow specular reflections with a given
probability~\cite{succi02,bib:tang-tao-he-2005,bib:sbragaglia-succi-2005},
or to apply diffuse
scattering~\cite{bib:ansumaili-karlin-2002,bib:sofonea-sekerka-2005,bib:niu-shu-chew-2004}.
The method we apply here follows the latter idea and uses a second order
accurate on-site fixed velocity boundary condition to simulate wall slippage.
The on site velocity boundary condition is used to set a required slip length
on the patterned surface. For the details of the implementation we refer the
reader to~\cite{HechtHarting2010,ahmed.nk:2009}.

Our geometry of configuration is the same as sketched in
Figure~\ref{fig:geometry}, but in simulations we employ periodic boundary
conditions in $x$ and $z$-direction, which allows to reduce the simulation
domain to a pseudo-2D system. To drive the flow a constant pressure
gradient is applied in $x$ direction by means of a homogeneous
acceleration, $g$, in the whole fluid domain.  Even though the simulation
domain can be reduced to be pseudo-2D, we find that the simulation of
fluid flow with a large slip in the thin channel limit still requires a
system of several million cells in order to properly resolve the velocity
field. The key issue is here that in order to resolve large slip lengths a
minimum channel height is necessary. To reach the thin channel limit the
stripe length has to be increased to even larger values.

Following the definition given for the Fourier analysis in
Sec.~\ref{fourier},
all heights $H$ and slip-lengths $b$ are given non dimensionalised for a
stripe length of $L=2\pi$. The resolution of the simulated system is then
given by the lattice constant
\begin{equation}
\Delta x = \frac{H L}{2\pi\mathcal{N}},
\end{equation}
where $\mathcal{N}$ is the number of discretization points used to resolve
the height of the channel. While systems with a height of $H=1.0$ can be
simulated using a discretization of $1\times32\times200\Delta x^3$ only,
decreasing $H$ to $0.1$ causes the required system size to be
$1\times70\times4400\Delta x^3$. To successfully recover the exact results
in the thin channel limit ($H=0.01$) a system of size
$1\times104\times64000\Delta x^3$, has to be simulated.

The number of timesteps required to reach a steady state depends on the channel
height, the velocity of the flow as determined by the driving acceleration as
well as the fraction of slip and no slip area at the surface. For the
simulations conducted in the thin channel limit a steady state velocity field
exactly fitting the theoretical prediction develops after one to four million
timesteps. In the thick channel limit, however, the number of timesteps
required can be an order of magnitude larger limiting the maximum feasible
system height. Moreover, the transition between slip and no slip stripes
induces a distortion of the flow field with a range of $\simeq 3 \Delta x$. In
order to keep the induced error below an acceptable limit, a minimum resolution
of the channel length of $64 \Delta x$ is maintained. Additionally, the
maximum flow velocity is limited due to the low Mach number assumption of our
lattice Boltzmann implementation. In effect, the acceleration modelling the
pressure gradient has to be reduced increasing the time required for
convergence. For example, a simulation domain of $1\times1024\times64 \Delta
x^3$ as used to model the thick channel limit at $H\simeq 100$ requires 12.5 million
timesteps to equilibrate.

\begin{figure}[h]
   \includegraphics[ width=6.5 cm, clip]{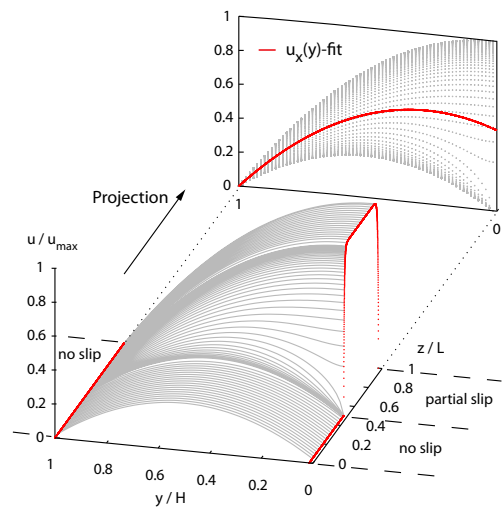}
	\caption{Illustration of the measurement of effective slip eigenvalues
from lattice Boltzmann simulations. The velocity information of the whole
domain is projected onto a single plane. To this cloud of data,
Eq.~(\ref{eq:vFit}) is fitted, effectively averaging the flow field over the
whole channel.}
   \label{fig:BeffMeasure}
 \end{figure}

We compare measurements of $b_{\rm eff}$ by permeability estimates and velocity
profiles, respectively. The permeability is calculated from measurements of the
flow rate, according to Eq.~(\ref{Darcy_tensor}). This allows to determine the
effective slip by Eq.~(\ref{downstream_b}).  Alternatively, the profiles of the
velocity in flow direction are averaged over the whole system by projecting the
velocity information of the whole domain onto a single plane. Then, the
effective slip length $b_{\rm eff}$ is found by a Levenberg-Marquardt fit of
the adjusted Hagen-Poiseuille equation
\begin{equation}\label{eq:vFit}
u_{x}(y)=\frac{g H^2}{2 \eta} \left(\frac{(y-H)^2}{H^2}-\frac{(y-H) + b_{\rm eff}}{(H+b_{\rm eff})} \right),
\end{equation}
with the known $g$ and $H$ (see Fig.~\ref{fig:BeffMeasure} for
an illustration).

The error of the effective slip measurements is determined by two factors,
namely the resolution of the channel height and the absolute slip length
of the partially slipping stripes. For poorly resolved channels with $10 <
\mathcal{N} < 30$ and small slip lengths the permeability measurements
still produce accurate results, whereas a fit of the velocity profiles
fails. If $b \gg H$, however, the quality of the obtained data declines.
For an increase in resolution ($30 < \mathcal{N} \le 100$) both approaches
allow very precise measurements for intermediate slip lengths of up to two
orders of magnitude larger than the channel height. However, if $b$ is
increased further, due to discretization the error in the effective slip determined by the
permeability measurement increases significantly rendering this method
inefficient since a higher resolution would be required. For example, to
keep the error in the determination of a prescribed slip in the order of
$10^5\Delta x$ below 5\% the permeability method requires the channel
height to be resolved by 200 lattice sites, while for the measurement by
fitting the velocity profiles 100 sites suffice.

To validate the concept of a tensorial slip by simulations of a flow past
tilted stripes, we do not rotate the surface pattern with respect to the
lattice, but instead change the direction of the acceleration in the
$y-z$-plane. This avoids
discretization errors due to the underlying regular lattice occurring in case
of a rotated surface pattern. We extract the
downstream slip by projecting the slip measured on the main axes onto
the pressure gradient direction.

\section{Results and Discussion}

\begin{figure}[h]
  \includegraphics[width=6.5cm,clip]{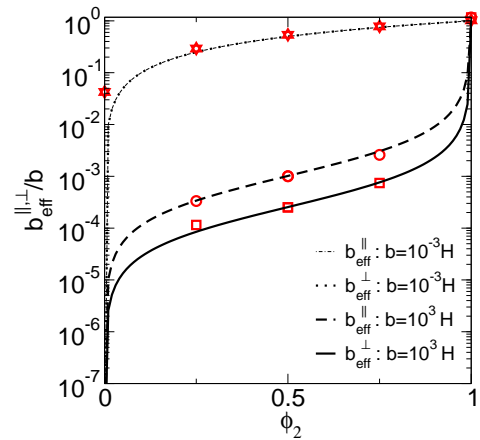}
  \caption{Eigenvalues of the effective slip length tensors simulated in the
limit of a thin channel (symbols). The lines represent results of theoretical
calculations by Eqs.~(\ref{beff_longit_smallH1}) and
(\ref{beff_longit_smallH}). The data show that at small $b$ the eigenvalues of
$\textbf{b}_{\rm eff}$ decrease as compared to a large local slip at the gas
sector, and that the slip-length tensor becomes isotropic resulting in $b_{\rm
eff}^{\perp, \parallel}$ to become hardly distinguishable in the $b=10^{-3}H$
case.}
  \label{fig:ComparesmallH}
\end{figure}

In this section we compare results of our LB simulation with  analytical example calculations and numerical solutions of
the dual series (\ref{DoubleSeries_a}), (\ref{DoubleSeries_b}) and
(\ref{DoubleSeries_c}), (\ref{DoubleSeries_d}) (see Appendix~\ref{A1}).

As a benchmark for the simulation, we start with a thin channel, where
striped surfaces were shown to provide rigorous upper and lower Wiener
bounds on the effective slip over all possible two-phase
patterns~\cite{feuillebois.f:2009}. In order to reach the thin channel limit, a
dimensionless height of $H=0.01$ is chosen.  The slip lengths are set to
$b=10^{-3}H$ ($0.1 \Delta x$) and $b=10^{3}H$ ($102000 \Delta x$),
differing each three orders of magnitude from the channel height and
reaching the limits of small- (cf. Eqs.~(\ref{xx1}) and (\ref{xx2})) as
well as large slip (Eqs.~(\ref{yy1}) and (\ref{yy2})). As preparatory
tests showed, a minimum channel height of $H=100 \Delta x$ is required to
measure slip-lengths of $b=10^{3}H$ corresponding to $10^5\Delta x$. For
a dimensionless height of $H=0.01$, we choose a simulation domain of
$1\times104\times64000 \Delta x^{3}$.  For each of the two slip-lengths, longitudinal and transverse flow was simulated for a different fraction of surface gas phase, ranging from no-slip ($\phi_{2}=0$) to homogeneous partial
slip ($\phi_{2}=1$). The local acceleration here (and below for thin channel simulations) was kept at $g=10^{-6}\Delta x/\Delta t^2$.

Fig.~\ref{fig:ComparesmallH} shows
the exact eigenvalues of the effective slip tensor in the thin channel
limit, Eqs.~(\ref{beff_longit_smallH1}), (\ref{beff_longit_smallH}), for
both slip lengths $b$.  The fit
of the simulation data and the analytical limits is excellent for all
separations. In the case of small local slip in the thin channel the
effective slip remains isotropic despite of the inhomogeneity of the
boundary. For large local slip, we observe truly tensorial effective slip
and highly anisotropic flow over the surface. These simulations demonstrate that finite size effects and resolution
effects are well controlled, and the size of the system is sufficient
to avoid artifacts. Another important point to note is that in our
theoretical analysis all equations were derived ignoring stripe edge
effects. An excellent agreement between theoretical and simulation
results indicates that the edge effects do not influence the simulation
results significantly.

\begin{figure}[h]
   \includegraphics[width=6.5cm,clip]{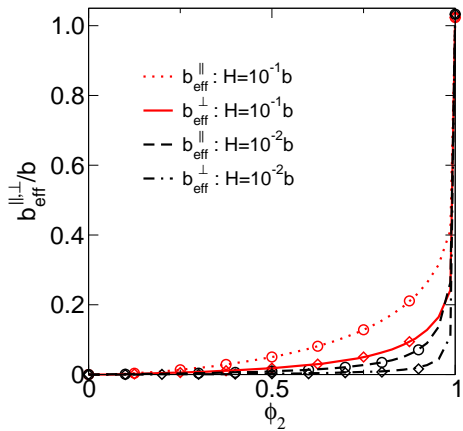}
	\caption{Simulation results of eigenvalues of $\textbf{b}_{\rm eff}$
as a function of fraction of gas sectors, $\phi_{2}$, in the limit of thick channel
(symbols). Lines represent corresponding theoretical values obtained by a
numerical solution of (\ref{DoubleSeries_a}), (\ref{DoubleSeries_b}) and
(\ref{DoubleSeries_c}), (\ref{DoubleSeries_d}).}
   \label{fig:Beffnum}
 \end{figure}

Fig.~\ref{fig:Beffnum} shows the eigenvalues of the effective slip length
tensor as a function of $\phi_{2}$ for a thick gap.  For these simulations the acceleration has been reduced down $g=10^{-7}\Delta x/\Delta t^2$ to obey the low Mach number (see Section \ref{sec:lbm})
limit. The time to reach a stable state increased then to
$15\cdot10^6\Delta t$. Simulation results are presented
for two different slip lengths of $b = 1.0$ and $b = 10.0$ in a system
of $H=0.1$, where $L$ is now resolved by $4400$ lattice sites. The theoretical solutions represented by the lines were obtained by the dual series
approach.
We find that the fit is excellent for all fractions of the slipping area, indicating
that our semi-analytical theory is extremely accurate. The data presented in Fig.~\ref{fig:Beffnum} show larger effective slip for a lower
slip to height ratio, i.e. a thicker channel. This illustrates well the
earlier suggestion that effective boundary conditions for this channel geometry are controlled
by the smallest length scale of the problem~\cite{belyaev.av:2010b}.

\begin{figure}
  \includegraphics[width=6.5cm,clip]{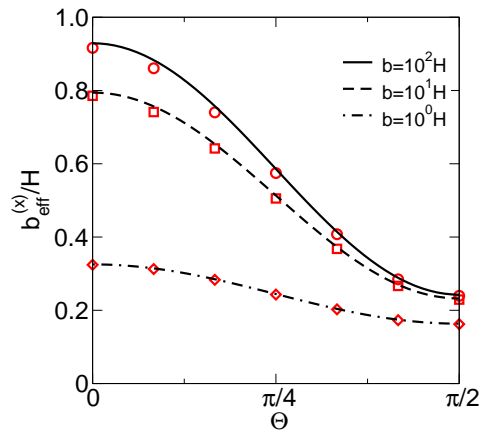}
  \includegraphics[width=6.5cm,clip]{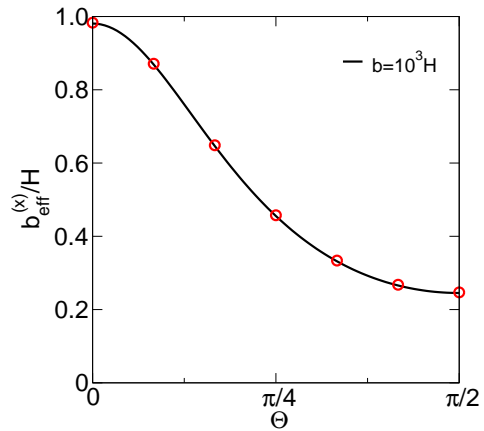}
   \caption{Downstream effective slip lengths of the thin channel
     simulated at $\phi_{2}=0.5$ for stripes inclined at different
     angle $\Theta$ (symbols). Especially in the limiting case (b)
     the value of $b^{(x)}_{\rm eff}$ has to be calculated from the
     measured eigenvalues by Eq.(\ref{b_eff_x_1}). All the lines are
     predicted theoretically downstream slip lengths: (a) Calculated
     by using Eq.(\ref{beff_def1}) [or Eq.(\ref{b_eff_x_1})] with
     eigenvalues of $\textbf{b}_{\rm eff}$ determined from numerical
     solutions of (\ref{DoubleSeries_a}),(\ref{DoubleSeries_b}) and
     (\ref{DoubleSeries_c}),(\ref{DoubleSeries_d}); (b) Calculated
     with Eq.({\ref{b_eff_x_thin}}).} \label{fig:BeffInflow}
 \end{figure}

\begin{figure}
   \includegraphics[width=6.5cm,clip]{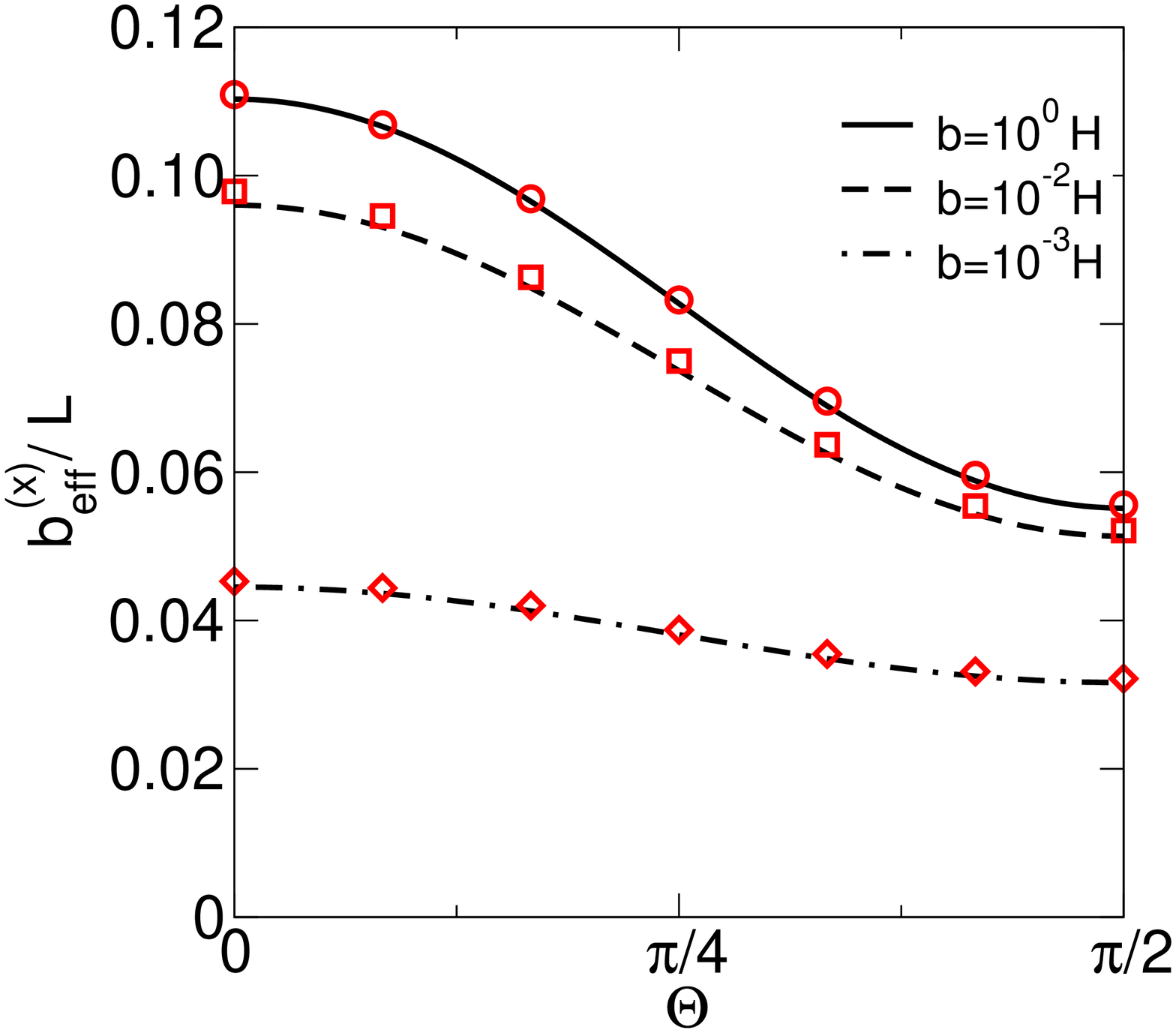}
   \includegraphics[width=6.5cm,clip]{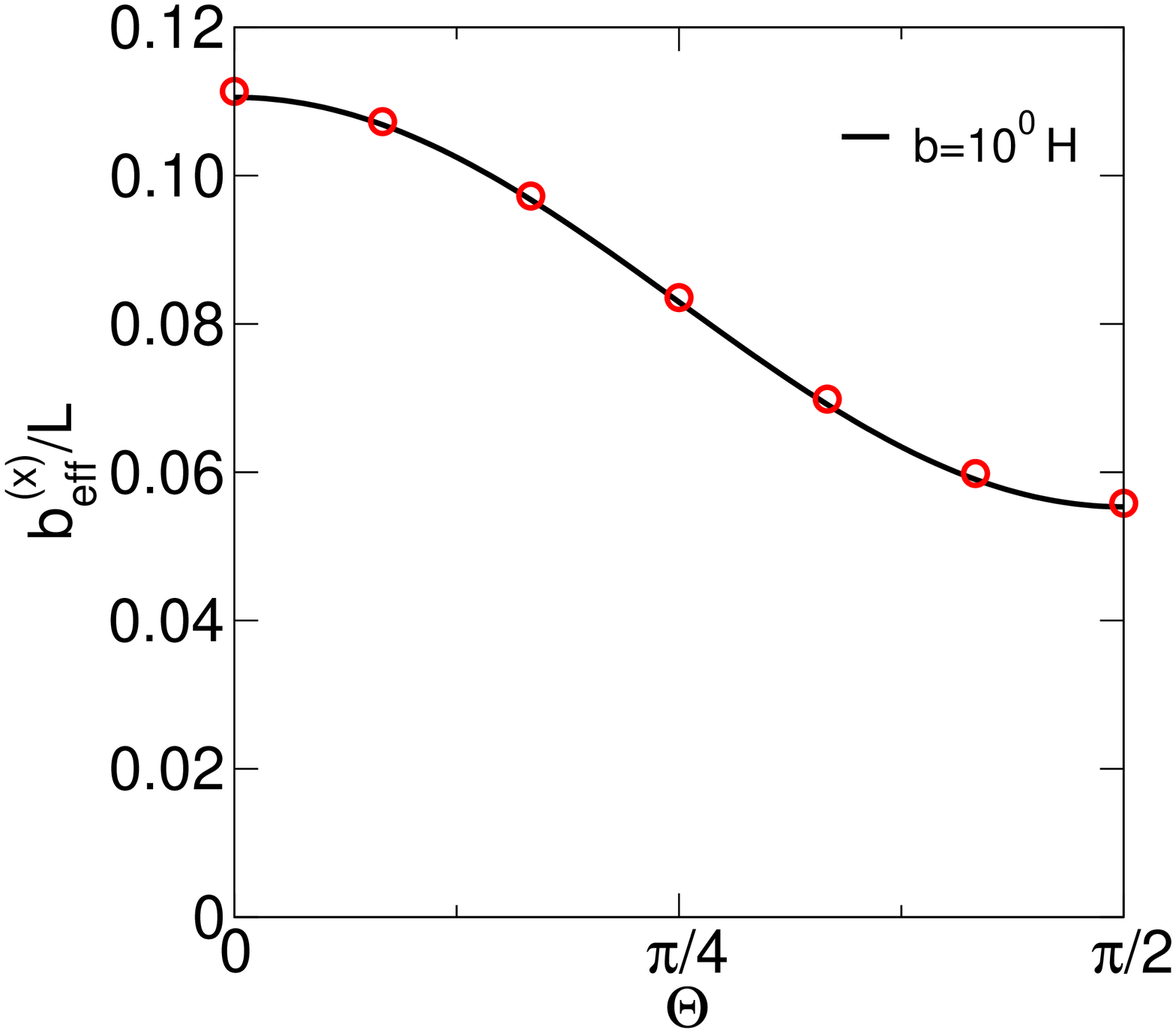}
   \caption{Effective downstream slip lengths for tilted stripes (the thick
channel limit) simulated at $\phi_{2}=0.5$ (symbols). All the lines are
predicted theoretically downstream slip lengths: (a) Calculated by using
Eq.(\ref{beff_def1}) [or Eq.(\ref{b_eff_x_1})] with eigenvalues of
$\textbf{b}_{\rm eff}$ determined from numerical solutions of
(\ref{DoubleSeries_a}), (\ref{DoubleSeries_b}) and
(\ref{DoubleSeries_c}), (\ref{DoubleSeries_d}); (b) Calculated with
Eq.~(\ref{b_eff_x_thick}) with eigenvalues evaluated with
Eqs.~(\ref{beff_ort_largeH_id}) and (\ref{beff_ort_largeH_id2}).}
\label{fig:BeffInflow2}
 \end{figure}

To check the validity of the tensorial slip approach, we now orient the
texture relative to the $x$-axis, which in our model is always aligned
with the applied pressure gradient. Fig.~\ref{fig:BeffInflow} and
\ref{fig:BeffInflow2} show two sets of effective downstream slip lengths
simulated with several $\Theta$, but fixed $H=0.1$ and $\phi_{2}=0.5$,
which results in a maximum transverse flow in a thin channel
situation~\cite{feuillebois.f:2010b}.

In the first set (Fig.~\ref{fig:BeffInflow}), we consider thin channels
and vary $b/H$ from 1 to 1000. Fig.~\ref{fig:BeffInflow}a shows simulation
data obtained using a channel of height $H=0.1$. Further, theoretical curves calculated
with Eq.~\ref{beff_def1} are presented. Here, eigenvalues of the slip-length tensor
are obtained by numerical solution of the dual series. The fits of the simulation data are in very good
agreement with the numerical solutions of the dual series suggesting the validity of
the concept of a tensorial slip in a thin channel situation. Note that the simulation results of Fig.~\ref{fig:BeffInflow}a cannot be compared with the analytical expression, Eq.(\ref{b_eff_x_thin}), because Fig.~\ref{fig:BeffInflow}a is based on a relatively moderate value of local slip at the gas sectors, whereas
Eq.(\ref{b_eff_x_thin}) requires very large $b$. To validate predictions of this analytical
formula, the channel height was decreased down to $H=0.01$. Simulation results are presented in
Fig.~\ref{fig:BeffInflow}b, confirming the surprising accuracy of a simple analytical expression, Eq.(\ref{b_eff_x_thin}). We remark that in this important limit of validity of Eq.(\ref{b_eff_x_thin}), $b^{(x)}_{\rm{eff}}/H$ is quite large,
although $b^{(x)}_{\rm{eff}}$ itself is small. This may have implications
for a reduction of a hydrodynamic drag
force~\cite{asmolov.es:2011,belyaev.av:2010b}.  Also passive mixing might
be an interesting application, since the anisotropy of flow is very large,
which is optimal for a transverse flow generation~\cite{feuillebois.f:2010b}. We suggest that our simple asymptotic result could be intensively used to simplify theoretical analysis of these important phenomena.

\begin{figure}
   \includegraphics[width=6.5cm,clip]{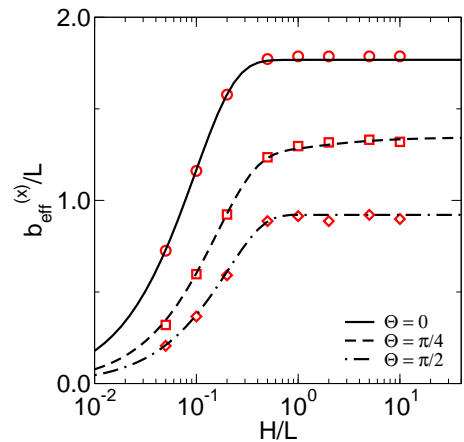}
   \caption{Effective downstream slip lengths at different channel thickness  ($\phi_2=0.75$, $b/L=5.0$). Symbols illustrate the simulation data, and curves show theoretical predictions.}
\label{fig:BeffvsH}
 \end{figure}

In the second set as shown in Fig.~\ref{fig:BeffInflow2}  a thick channel
(of height $H=100$) is simulated.  Fig.~\ref{fig:BeffInflow2}a plots simulation
results for several $b/H$ varying from $10^{-3}$ to 1(symbols). Similarly to previous examples, we found a very good agreement between simulation results and predictions of tensorial Eq.(\ref{beff_def1}) with eigenvalues computed with our semi-analytical theory. We found that already for the case
$b=H=100$ our simulations reach the limit of large slip in the thick channel, so that the comparison with analytical solutions is possible. To examine this more
closely, the simulations results obtained in this limit are reproduced in
Fig.~\ref{fig:BeffInflow2}b. Also included are the theoretical results calculated with asymptotic formulae, Eqs.(\ref{beff_ort_largeH_id}) and (\ref{beff_ort_largeH_id2}), which perfectly fit simulation data.

Finally, we simulate the downstream slip length as a function of the channel thickness with the focus on the intermediate gap situation. Fig.~\ref{fig:BeffvsH} shows the typical simulation results (the example corresponds
to $b/L = 5.0$ and $\phi_2 = 0.75$) and demonstrates that the
effective slip lengths increase with $H$ and saturate for a thick
gap. This fully confirms the statement that an effective boundary condition
is not a characteristic of the liquid-solid interface alone, but
depends on the flow configuration and interplay between the
typical length scales of the problem. Again, the simulation and theoretical data are in the excellent agreement for longitudinal, transverse, and tilted stripes. Thus, Fig.~\ref{fig:BeffvsH} unambiguously shows that the tensorial slip boundary condition, originally justified for a thick
channel, to any channel thickness can be generalized to any channel thickness.

\section{Conclusion}
We have investigated pressure-driven flow in a flat-parallel channel with one
hydrophilic and one super-hydrophobic surface, and have given some general
theoretical arguments showing that a concept of an effective tensorial slip is
valid for any thickness (compared to a super-hydrophobic texture scale). The
eigenvalues of the effective slip-length tensor depend on the gap, so that they
cannot be viewed as a local property of the super-hydrophobic surface, except
in the thick channel limit. Instead, the slip-length tensor represents a global
characteristic of the channel. The mathematical properties of the slip-length
and permeability tensors allowed us to derive a simple analytical formula for
an effective downstream slip length in case of inclined to a pressure gradient
textures. Our analysis is validated by means of LB simulations.

\begin{acknowledgments}
We have benefited from discussions with M.Z.~Bazant at the initial stage of
this study. This research was partly supported by the Russian Academy of
Science (RAS) through its priority program `Assembly and Investigation of
Macromolecular Structures of New Generations', by the Netherlands Organization
for Scientific Research (NWO/STW VIDI), and by the German
Science Foundation (DFG) through its priority program `Micro- and nanofluidics'
and the collaborative research center (SFB) 716. We acknowledge computing
resources from the J\"ulich Supercomputing Center and the Scientific
Supercomputing Center Karlsruhe.
\end{acknowledgments}

\appendix
\section{Numerical method}\label{A1}

Equations (\ref{DoubleSeries_a}), (\ref{DoubleSeries_b}) and
(\ref{DoubleSeries_c}), (\ref{DoubleSeries_d}) provide a complete
description of hydrodynamic flow and effective slip in eigendirections.
Their exact solution is possible for the limits of a thin and a thick
channel only. In order to solve the problem for general channel thickness,
the following numerical algorithm has been used.

It is convenient to change to dimensionless values. We, therefore, choose
$L/(2\pi)$ as the reference length scale and $\sigma H L/(4\pi\eta)$ as
the velocity scale. We make the substitution
\begin{equation}
   (x,y,z)=\frac{L}{2\pi}(\tilde{x},\tilde{y},\tilde{z}),\quad H=\frac{L}{2\pi}\tilde{H},\quad b=\frac{L}{2\pi}\tilde{b},
\end{equation}
\begin{equation}\label{scale}
   a_n = \frac{\sigma H L}{4\pi\eta}\tilde{a}_n, \quad n\geq 0,
\end{equation}
where non-dimensional variables are denoted by tildes.
This procedure gives the dual series problem for longitudinal flow in the form
\begin{eqnarray}\label{DoubleSeries_a_dimless}
   \tilde{a}_0\left(1+\frac{\tilde{b}}{\tilde{H}}\right)&+&\sum^\infty_{n=1} \tilde{a}_n \left[1+\tilde{b} n \coth(n \tilde{H}) \right] \cos(n \tilde{z})\nonumber\\&=&\tilde{b} ,\quad 0<\tilde{z}\leq c,   
\end{eqnarray}
\begin{equation}\label{DoubleSeries_b_dimless}
   \tilde{a}_0+\sum^\infty_{n=1} \tilde{a}_n \cos(n \tilde{z})=0,\quad c<\tilde{z}\leq \pi,
\end{equation}
where $c=\pi\phi_2=\pi\delta/L$. Similarly, the equations for the flow in
the direction orthogonal to the stripes is written as
\begin{eqnarray}\label{DoubleSeries_c_dimless}
   \tilde{a}_0\left(1+\frac{\tilde{b}}{\tilde{H}}\right)&+&\sum^\infty_{n=1} \tilde{a}_n \left[1+2 \tilde{b} n V(n \tilde{H}) \right] \cos(n \tilde{x})\nonumber\\&=&\tilde{b} ,\quad 0<\tilde{x}\leq c,   
\end{eqnarray}
\begin{equation}\label{DoubleSeries_d_dimless}
   \tilde{a}_0+\sum^\infty_{n=1} \tilde{a}_n \cos(n \tilde{x})=0,\quad c<\tilde{x}\leq \pi.
\end{equation}

After integrating Eq.~\ref{DoubleSeries_a_dimless} over $[0,\tilde{z}]$ and Eq.~\ref{DoubleSeries_c_dimless} over $[0,\tilde{x}]$),
we multiply the result by $\sin(m \tilde{z})$ ($\sin(m \tilde{x})$,
respectively), where $m$ is a nonnegative integer. We then integrate again
over $[0, c]$. Eq.~\ref{DoubleSeries_b_dimless} is multiplied by
$\cos(m \tilde{z})$ and Eq.~\ref{DoubleSeries_d_dimless} by $\cos(m
\tilde{x})$ and we then integrate over $[c, \pi]$. The resulting
equations are summarized to obtain a system of linear algebraic equations,
\begin{equation}\label{SLAE}
   \sum^\infty_{n=0} A_{nm} \tilde{a}_n= B_m,
\end{equation}
which can be solved with respect to $\tilde{a}_n$ by standard numerical algebra tools.

The coefficients for the longitudinal case are ($m\geq 0$)
\begin{equation}
   A^{\parallel}_{0m} = \left(1+\frac{\tilde{b}}{\tilde{H}}\right)\int\limits_0^c{\tilde{z} \sin(m \tilde{z}) d\tilde{z}} +\int\limits_c^\pi{\cos(m \tilde{z}) d\tilde{z}},
\end{equation}
\begin{eqnarray}
   A^{\parallel}_{nm} &=& \frac{1+\tilde{b} n \coth(n \tilde{H})}{n}
\int\limits_0^c{\sin(m \tilde{z}) \sin(n \tilde{z}) d\tilde{z}}
\nonumber\\&+& \int\limits_c^\pi{\cos(m \tilde{z})\cos(n \tilde{z})
d\tilde{z}},\quad n\geq 1,
\end{eqnarray}
\begin{equation}
   B^{\parallel}_m = \tilde{b} \int\limits_0^c{\tilde{z} \sin(m\tilde{z}) d\tilde{z}}.
\end{equation}
For transverse flow we get
\begin{eqnarray}
   A^{\perp}_{nm} &=& \frac{1+2 \tilde{b} n V(n \tilde{H})}{n} \int\limits_0^c{\sin(m \tilde{z}) \sin(n \tilde{z}) d\tilde{z}} \nonumber\\&+& \int\limits_c^\pi{\cos(m \tilde{z})\cos(n \tilde{z}) d\tilde{z}},\quad n\geq 1
\end{eqnarray}
\begin{equation}
   A^{\perp}_{0m} = A^{\parallel}_{0m}, \quad B^{\perp}_m=B^{\parallel}_m.
\end{equation}
~\\
For the numerical evaluation, the linear system is truncated and reduced
to a $N\times N$ matrix and the solution is then found to converge upon
truncation refinement. According to the definition
\begin{equation} \label{definition2}
   b_{\rm eff}=\frac{\langle u_s\rangle}{\left\langle \left(\frac{\partial u}{\partial y}\right)_s \right\rangle},
\end{equation}
the dimensionless and dimensional effective slip lengths are given by
\begin{equation}
   \tilde{b}_{\rm eff}=\frac{\tilde{a}_0}{1-\tilde{a}_0/\tilde{H}},
\end{equation}
\begin{equation}
   b_{\rm eff}=\frac{L}{2\pi}\tilde{b}_{\rm eff}.
\end{equation}


\begin{thebibliography}{10}
\bibitem{notice1}
Note that $a_n$ scale with $\sigma H/\eta$. By using an appropriate scale for coefficients $a_n$ (see Eq.(\ref{scale})) it is easy to show that the r.h.s. of dual series does not scale with $H$.
\end{thebibliography}

\end{document}